\documentclass[entropy,article,accept,moreauthors,10pt,a4paper]{mdpi}

\firstpage{1}
\pubvolume{1}
\issuenum{1}
\articlenumber{0}
\pubyear{2025}
\copyrightyear{2025}
\externaleditor{Gianpiero Colonn}
\datereceived{17 November 2025}
\daterevised{5 December 2025} 
\dateaccepted{9 December 2025}
\datepublished{ }
\hreflink{https://doi.org/} 

\newcommand{\dd}{\mathrm{d}}
\newcommand{\DD}{\mathrm{D}}
\newcommand{\FF}{\mathrm{F}}
\newcommand{\HH}{\mathrm{H}}
\newcommand{\ee}{\mathrm{e}}

\newcommand{\TB}{T_{\text{B}}^*}

\newcommand\beq{\begin{equation}}
\newcommand\eeq{\end{equation}}
\newcommand\beqa{\begin{eqnarray}}
\newcommand\eeqa{\end{eqnarray}}
\newcommand{\nn}{\nonumber\\}
\def\bal#1\eal{\begin{align}#1\end{align}}

\Title{{On} the First Quantum Correction to the Second Virial Coefficient of a Generalized Lennard-Jones Fluid}

\TitleCitation{On the First Quantum Correction to the Second Virial Coefficient of a Generalized Lennard-Jones Fluid}

\Author{{{Daniel}  Parejo} $^{1}$ and {Andr\'es Santos} $^{1,2,}$*\orcidC{}}

\AuthorNames{Daniel Parejo and Andr\'es Santos}

\AuthorCitation{Parejo, D.; Santos, A.}

\address{%
$^{1}$ \quad Departamento de F\'isica, Universidad de Extremadura, {E-06006} Badajoz, Spain; {dparejof@alumnos.unex.es} \\
$^{2}$ \quad Instituto de Computaci\'on Cient\'ifica Avanzada (ICCAEx), Universidad de
Extremadura, \mbox{E-06006 Badajoz, Spain}}

\corres{Correspondence: andres@unex.es}

\abstract{
We derive an explicit analytic expression for the first quantum correction to the second virial coefficient of a $d$-dimensional fluid whose particles interact via the generalized Lennard-Jones  $(2n,n)$ potential.
By introducing an appropriate change of variable, the correction term is reduced to a single integral that can be evaluated in closed form in terms of parabolic cylinder or generalized Hermite functions.
The resulting expression compactly incorporates both dimensionality and stiffness, providing direct access to the low- and high-temperature asymptotic regimes.
In the special case of the standard Lennard-Jones fluid ($d=3$, $n=6$), the formula obtained is considerably more compact than previously reported representations based on hypergeometric functions.
The knowledge of this correction allows us to determine the first quantum contribution to the Boyle temperature, whose dependence on dimensionality and stiffness is explicitly analyzed, and enables quantitative assessment of quantum effects in noble gases such as helium, neon, and argon.
Moreover, the same methodology can be systematically extended to obtain higher-order quantum corrections.
}


\begin{document}


\section{Introduction}
In a semiclassical fluid, the~second virial coefficient can be expressed {as follows}~\cite{K55,KMS55,DW62,M66,BLK69,B71,ME16,ZGCPBEHANLL25}:
\beq
\label{0}
B_2(T)=B^{(0)}(T)+\frac{\hbar^2}{m}B^{(1)}(T)+\mathcal{O}\left(\frac{\hbar^4}{m^2}\right),
\eeq
where
\beq
\label{1}
B^{(0)}(T)=\frac{\Omega_d}{2}\int_0^\infty\dd r\, r^{d-1} \left[1-\ee^{-\beta\phi(r)}\right],\quad \Omega_d=\frac{2\pi^{\tfrac{d}{2}}}{\Gamma(\tfrac{d}{2})},
\eeq
is the classical contribution, and~\beq
\label{2}
B^{(1)}(T)=\frac{\Omega_d\beta^3}{24}\int_0^\infty \dd r\, r^{d-1} \ee^{-\beta\phi(r)}\left[\frac{\dd\phi(r)}{\dd r}\right]^2
\eeq
represents the first quantum correction. In~Equations~\eqref{1} and \eqref{2}, $\phi(r)$ denotes the pair potential, $\beta\equiv (k_BT)^{-1}$, and~$\Omega_d$ is the total solid angle in $d$ dimensions.

{The prototypical pair potential in liquid-state theory is the Lennard-Jones (LJ) potential, written as follows:}
\beq
\label{3}
\phi(r)=4\epsilon\left[\left(\frac{\sigma}{r}\right)^{2n}-\left(\frac{\sigma}{r}\right)^n\right],
\eeq
where $\epsilon$ and $\sigma$ set the energy and length scales, respectively. The~standard LJ (sLJ) fluid corresponds to $d=3$ and $n=6$,
while the generalized Lennard-Jones (gLJ) model allows arbitrary dimensionality $d$ and stiffness parameter $n>d$. In~the limit $n\to\infty$, the~gLJ potential of Equation~\eqref{3} approaches the hard-sphere~potential.

By introducing the reduced (dimensionless) coefficients, written as follows:
\beq
B_c\equiv \frac{2d}{\Omega_d\sigma^d}B^{(0)},\quad B_q\equiv \frac{24\epsilon}{\Omega_d\sigma^{d-2}}B^{(1)},
\eeq
{we}
 obtain
\begin{subequations}
\beq
B_c(T^*)=d \int_0^\infty \dd x\, x^{d-1} \left[1-\ee^{-4\beta^*(x^{-2n}-x^{-n})}\right],
\eeq
\beq
\label{4a}
B_q(T^*)=16n^2{\beta^*}^3 \int_0^\infty \dd x\, x^{-(2n+3-d)} \ee^{-4\beta^*(x^{-2n}-x^{-n})}\left(1-4x^{-n}+4x^{-2n}\right),
\eeq
\end{subequations}
{where}  $T^*\equiv 1/\beta^*=k_BT/\epsilon$ is the reduced~temperature.

Several equivalent representations of $B_c$ for the sLJ fluid can be found in the literature (see, for~instance, Ref.~\cite{ZGCPBEHANLL25} and references therein).
Perhaps the most compact expression---valid for the gLJ fluid---is {Section 3.7}~in~\cite{S16}:
\beq
\label{5}
B_c(T^*)=\Gamma(1-\tfrac{d}{n})\left(8\beta^*\right)^{\tfrac{d}{2n}}\ee^{\beta^*/2}\DD_{\frac{d}{n}}\left(-\sqrt{2\beta^*}\right),
\eeq
where we use the integral representation
\beq
\label{6}
\DD_a(z)=\frac{\ee^{-z^2/4}}{\Gamma(-a)}\int_0^\infty \dd t\, t^{-a-1}\left[\ee^{-t^2/2-zt}-\Theta(a)\right],\quad a<1,
\eeq
for the parabolic cylinder function {(Equation~(12.5.1)}  Available online:  \url{https://dlmf.nist.gov/12.5.E1}  {(accessed on 16 November 2025)})~\cite{DLMF}.
In Equation~\eqref{6}, $\Theta(a)$ denotes the Heaviside step function. It  ensures convergence of the integral at the lower integration limit. For~$a < 0$, the~integrand $t^{-a-1}\ee^{-t^2/2-zt}$ is integrable at $t = 0$ without modification. However, for~$0 < a < 1$, the~behavior $t^{-a-1}$ near $t = 0$ causes a divergence. The~subtraction of $\Theta(a) = 1$ removes the leading constant term from the small-$t$ expansion of the exponential, regularizing the integral while preserving the correct value of $\DD_a(z)$ through analytic continuation in $a$.

Naturally, the~situation is more involved for the quantum contribution $B_q$. In~a recent work, Zhao~et~al.~\cite{ZGCPBEHANLL25} derived a linear, second-order homogeneous ordinary differential equation for the sLJ coefficient $B_q$. From~its solution, they obtained the following:
\beq
\label{7}
B_q(T^*)=\frac{1}{3\times 2^{\tfrac{11}{6}}}\left[\Gamma(\tfrac{5}{12}) F(T^*)-\Gamma(-\tfrac{1}{12}) G(T^*)\right],
\eeq
where
\begin{subequations}
\label{8}
\beq
F(T^*)={\beta^*}^{\tfrac{19}{12}}\left[72\, _1\FF_1(\tfrac{5}{12};\tfrac{1}{2};\beta^*)-22\, _1\FF_1(\tfrac{5}{12};\tfrac{3}{2};\beta^*)\right],
\eeq
\beq
G(T^*)={\beta^*}^{\tfrac{13}{12}}\left[12\beta^*\, _1\FF_1(\tfrac{11}{12};\tfrac{3}{2};\beta^*)+11\, _1\FF_1(-\tfrac{1}{12};\tfrac{1}{2};\beta^*)\right].
\eeq
\end{subequations}
{Here,}   $_1\FF_1\left(a;b;z\right)$ denotes the Kummer confluent hypergeometric function {(Equation}~(13.2.2)  Available online: \url{https://dlmf.nist.gov/13.2.E2}  {(accessed on 16 November 2025)})~\cite{DLMF}.
The result given by Equations~\eqref{7} and \eqref{8} was first obtained by Michels~\cite{M66}.

\section{First Quantum Correction to the Second Virial~Coefficient}
Our goal is to derive an alternative, more compact expression for $B_q$ in the broader case of the gLJ fluid.
We begin by stating the final result:
\beq
\label{9}
B_q(T^*)=n\frac{\Gamma(2-\tfrac{d-2}{n})}{8}\left(8\beta^*\right)^{\frac{d-2}{2n}+1}\ee^{\beta^*/2}\left[\DD_{\frac{d-2}{n}}\left(-\sqrt{2\beta^*}\right)+\DD_{\frac{d-2}{n}-2}\left(-\sqrt{2\beta^*}\right)\right].
\eeq

Before proving Equation~\eqref{9}, we list several useful properties of the parabolic cylinder function~\cite{DLMF}:
\begin{subequations}
\beq
\label{10}
\DD_a(z)=z\DD_{a-1}(z)+(1-a)\DD_{a-2}(z),
\eeq
\beq
\label{10a}
\frac{\partial \DD_a(z)}{\partial z}=a\DD_{a-1}(z)-\frac{z}{2}\DD_a(z),
\eeq
\beq
\label{10b}
\lim_{z\to 0}\DD_a(z)=\frac{\sqrt{\pi}2^{\tfrac{a}{2}}}{\Gamma(\frac{1-a}{2})},\quad
\lim_{z\to \infty}\DD_a(-z)=\frac{\sqrt{2\pi}}{\Gamma(-a)}\ee^{z^2/4}z^{-a-1},
\eeq
\beq
\label{10c}
\DD_0(z)=\ee^{-z^2/4},\quad \DD_{-2}(z)=\ee^{-z^2/4}-\sqrt{\frac{\pi}{2}}\ee^{z^2/4}z\,\mathrm{erfc}\left(\frac{z}{\sqrt{2}}\right),
\eeq
\beq
\label{10d}
\DD_a(\sqrt{2} z)=2^{-\tfrac{a}{2}}\ee^{-z^2/2}\HH_a(z).
\eeq
\end{subequations}
{Equation}~\eqref{10d} defines the generalized Hermite functions $\HH_a(z)$ for arbitrary (noninteger) degree $a<1$ ({Equation}~(12.7.2)  Available online:  \url{https://dlmf.nist.gov/12.7.E2}  {(accessed on 16 November 2025)})~\cite{DLMF}.

By introducing the change of variable $x\to t=\sqrt{8\beta^*}x^{-n}$ in Equation~\eqref{4a}, we obtain the following:
\beq
\label{11}
B_q(T^*)=\frac{n}{8}\left(8\beta^*\right)^{\tfrac{a}{2}+1}\int_0^\infty \dd t\, t^{-a+1} \ee^{-t^2/2-zt}\left(z^2+2zt+t^2\right),
\eeq
where we have used the  notation $a\equiv \tfrac{d-2}{n}$, $z\equiv -\sqrt{2\beta^*}$.
Using the {integral representation} of the parabolic cylinder function {[see Equation}~\eqref{6}{], Equation}~\eqref{11} can be rewritten as follows:
\bal
\label{12}
B_q(T^*)=&n\frac{\Gamma(2-a)}{8}\left(8\beta^*\right)^{\tfrac{a}{2}+1}\ee^{z^2/4}\Big[z^2\DD_{a-2}(z)
+2(2-a)z\DD_{a-3}(z)
\nn&
+(3-a)(2-a)\DD_{a-4}(z)\Big].
\eal
{This} expression is already quite compact, but~it can be further simplified. Iterative application of Equation~\eqref{10} yields the following:
\bal
\DD_a(z)=&z\left[z\DD_{a-2}(z)+(2-a)\DD_{a-3}(z)\right]+(1-a)\DD_{a-2}(z)\nn
=&z^2\DD_{a-2}(z)+(2-a)z\DD_{a-3}(z)+(2-a)\DD_{a-2}(z)-\DD_{a-2}(z).
\eal
{Next,} we apply Equation~\eqref{10} to the term $(2-a)\DD_{a-2}(z)$, which gives the following:
\beq
\DD_a(z)=z^2\DD_{a-2}(z)+2(2-a)z\DD_{a-3}(z)+(3-a)(2-a)\DD_{a-4}(z)-\DD_{a-2}(z).
\eeq
{Substituting} this identity into Equation~\eqref{12}, and~returning to the physical variables $a\to \tfrac{d-2}{n}$ and $z\to -\sqrt{2\beta^*}$, we recover Equation~\eqref{9}. In~terms of the generalized Hermite functions, Equation~\eqref{9} can be rewritten as follows:
\beq
\label{15}
B_q(T^*)=n\frac{\Gamma(2-\frac{d-2}{n})}{4}\left(4\beta^*\right)^{\frac{d-2}{2n}+1}\left[\HH_{\frac{d-2}{n}}\left(-\sqrt{\beta^*}\right)+2\HH_{\frac{d-2}{n}-2}\left(-\sqrt{\beta^*}\right)\right].
\eeq
{For} the particular case of the sLJ model ($d=3$, $n=6$), the following is written:
\bal
\label{16}
B_q(T^*)=&\frac{3\Gamma(\frac{11}{6})}{4}\left(8\beta^*\right)^{\frac{13}{12}}\ee^{\beta^*/2}\left[\DD_{\frac{1}{6}}\left(-\sqrt{2\beta^*}\right)+\DD_{-\frac{11}{6}}\left(-\sqrt{2\beta^*}\right)\right]\nn
=&\frac{3\Gamma(\frac{11}{6})}{2}\left(4\beta^*\right)^{\frac{13}{12}}\left[\HH_{\frac{1}{6}}\left(-\sqrt{\beta^*}\right)+2\HH_{-\frac{11}{6}}\left(-\sqrt{\beta^*}\right)\right].
\eal
{It} can be verified that Equation~\eqref{16} is equivalent to the combination of Equations~\eqref{7} and~\eqref{8}.

The limits given by Equation~\eqref{10b} allow us to determine the low- and high-temperature behaviors of $B_q(T^*)$ for the gLJ fluid:
\beq
\label{17}
\lim_{T^*\to 0}B_q(T^*)=n\sqrt{\pi}2^{\frac{d-2}{n}+1}\ee^{1/T^*}{T^*}^{-\frac{3}{2}},\quad \lim_{T^*\to \infty}B_q(T^*)=n\frac{\Gamma(2-\frac{d-2}{2n})}{2}\left(\frac{4}{T^*}\right)^{\frac{d-2}{2n}+1}.
\eeq
In the second equality of Equation~\eqref{17}, use has been made of the identity
$\Gamma(2x)/\Gamma(x)=2^{2x-1}\Gamma(x+\tfrac{1}{2})/\sqrt{\pi}$.

Interestingly, Equation~\eqref{9} simplifies considerably in the case of a two-dimensional fluid ($d=2$).
Using Equation~\eqref{10c}, we obtain the following:
\beq
\label{20}
B_q(T^*)=n\beta^*\left[2+\sqrt{\pi\beta^*}\ee^{\beta^*}\,\mathrm{erfc}(-\sqrt{\beta^*})\right].
\eeq
In this case, the~ratio $B_q/n$ is independent of the stiffness parameter $n$. It is worth mentioning that Equation~\eqref{20} also provides the ratio $B_q/n$  in the limit $n\to\infty$ for any~\mbox{dimensionality}.

\begin{figure}[H]
\includegraphics[width=6cm]{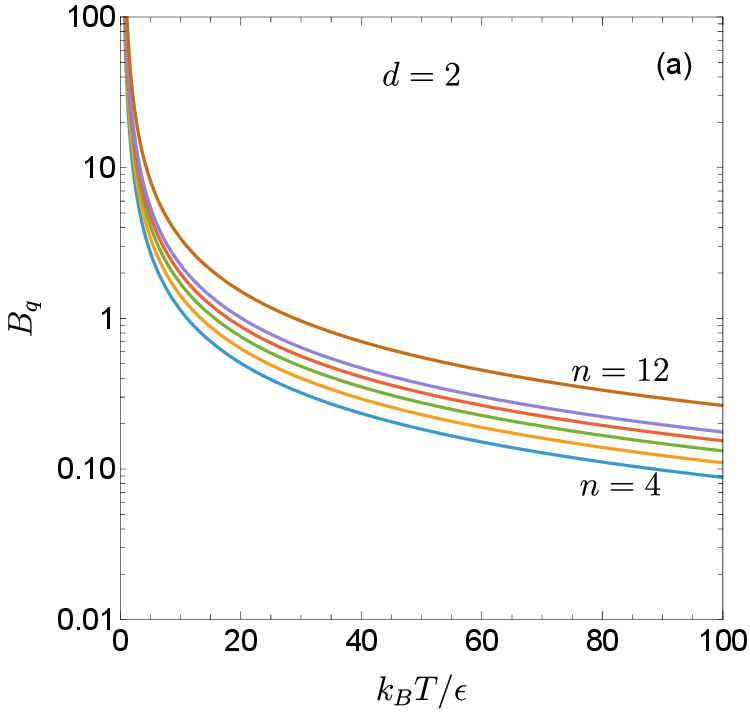}\includegraphics[width=6cm]{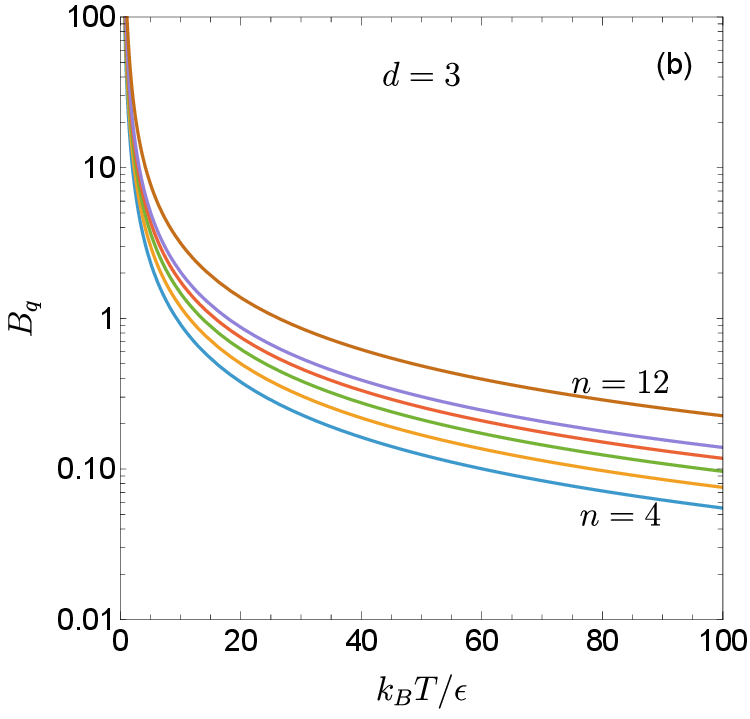}
%
\vspace{-3pt}
\caption{
{Reduced} first quantum correction to the second virial coefficient, as~given by Equation~\eqref{9}, for~the gLJ fluid with (\textbf{a}) $d=2$ and (\textbf{b}) $d=3$.
The curves correspond, from~bottom to top, to~stiffness parameters $n=4$, $5$, $6$, $7$, $8$, and~$12$.\label{fig1}}
\end{figure}

Figure~\ref{fig1} illustrates the temperature {dependence of} $B_q(T^*)$ for the two- and three-dimensional gLJ fluids with $n=4$, $5$, $6$, $7$, $8$, and~$12$.
As can be seen, for~a given $T^*$, the~reduced first quantum correction $B_q$ increases as the potential becomes stiffer.
Moreover, the~influence of $n$ is more pronounced at high temperatures than at low temperatures, consistent with the limiting behaviors described by Equation~\eqref{17}.
For the same values of $T^*$ and $n$, $B_q$ is larger in the two dimensions than in~three.

\section{First Quantum Correction to the Boyle~Temperature}

From Equation~\eqref{0}, the~reduced second virial coefficient of the gLJ fluid can be written~as follows:
\beq
\label{21}
\frac{2d}{\Omega_d\sigma^d}B_2(T^*)=B_c(T^*)+\frac{d}{12}qB_q(T^*)+\mathcal{O}(q^2),
\eeq
where the dimensionless quantum parameter
\beq
q\equiv \frac{\hbar^2}{m\sigma^2\epsilon}
\eeq
measures the expected magnitude of quantum~effects.

The Boyle temperature, $\TB$, is defined by the condition $B_2(\TB)=0$.
It marks the balance between the attractive and repulsive contributions to the intermolecular potential: the attractive interactions dominate for $T^*<\TB$, whereas the repulsive ones dominate for $T^*>\TB$.
In the semiclassical regime, the~Boyle temperature can be expanded as follows:
\beq
\label{22}
\TB=T_0^*-qT_1^*+\mathcal{O}(q^2),
\eeq
where $T_0^*$ is the classical Boyle temperature, i.e.,~the solution of $B_c(T_0^*)=0$, or~equivalently, $\DD_{d/n}(-\sqrt{2/T_0^*})=0$.
By inserting Equation~\eqref{22} into Equation~\eqref{21}, one obtains the first quantum correction to the Boyle temperature,
\beq
T_1^*=\frac{d}{12}\frac{B_q(T_0^*)}{\left.\partial B_c(T^*)/\partial T^*\right|_{T_0^*}}.
\eeq
Note that $\partial B_c/\partial T^*=-{T^*}^{-2}\partial B_c/\partial \beta^*$, where, from~Equations~\eqref{5} and~\eqref{10a}, {one finds}
\beq
\frac{\partial B_c}{\partial \beta^*}=\frac{d}{n}\left[\frac{B_c}{2\beta^*}-\frac{\Gamma(1-\frac{d}{n})}{\sqrt{2\beta^*}}\left(8\beta^*\right)^{\frac{d}{2n}}\ee^{\beta^*/2}\DD_{\frac{d}{n}-1}\left(-\sqrt{2\beta^*}\right)\right].
\eeq
Thus, one finally obtains the following:
\beq
\label{25}
T_1^*=n^2\frac{\Gamma(2-\frac{d-2}{n})}{3\Gamma(1-\frac{d}{n})}\left(\frac{T_0^*}{8}\right)^{\frac{1}{2}+\frac{1}{n}}
\frac{\DD_{\frac{d-2}{n}}(-\sqrt{2/T_0^*})+\DD_{\frac{d-2}{n}-2}(-\sqrt{2/T_0^*})}{\DD_{\frac{d}{n}-1}(-\sqrt{2/T_0^*})}.
\eeq

\begin{figure}[H]
\includegraphics[width=6cm]{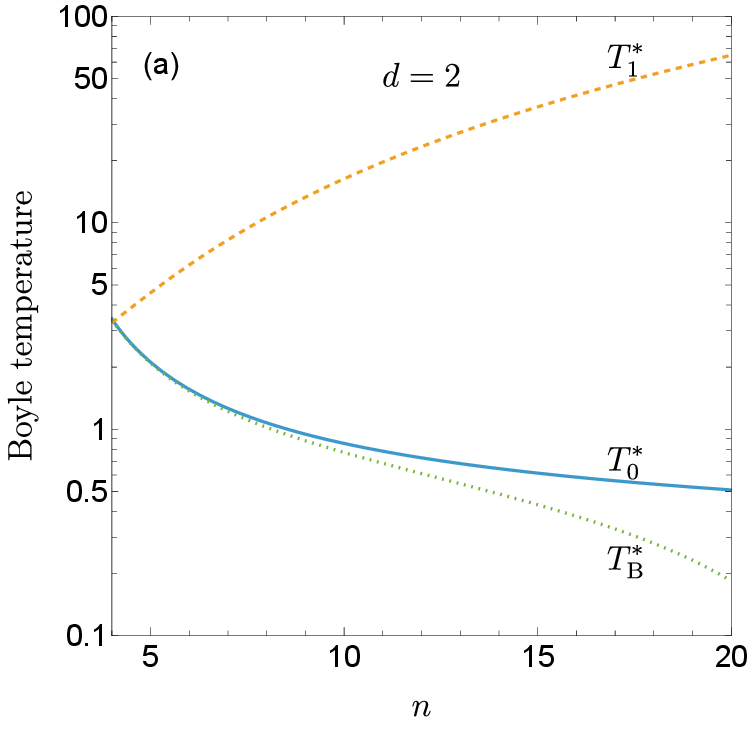}\includegraphics[width=6cm]{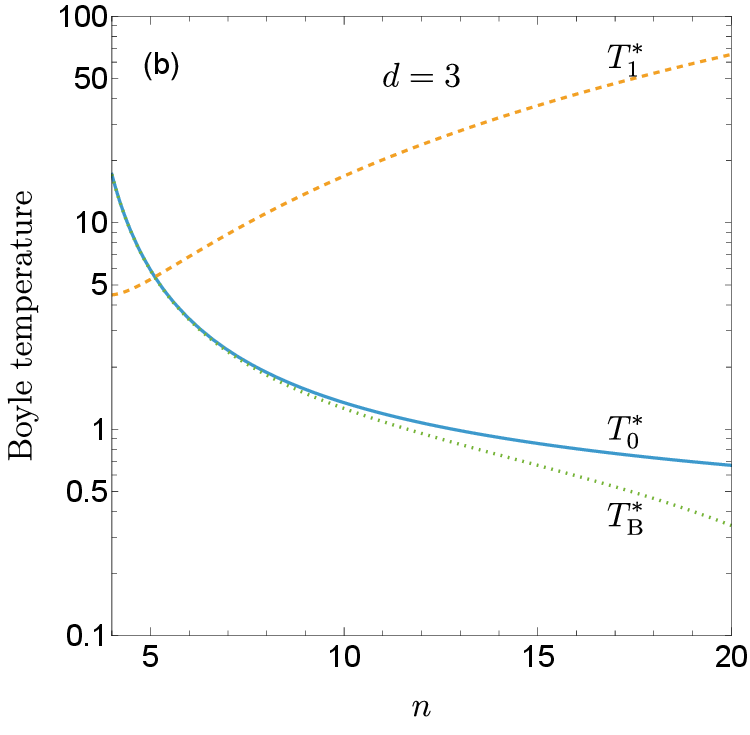}
%
\vspace{-3pt}
\caption{
Classical Boyle temperature $T_0^*$ (solid lines) and  its first quantum correction $T_1^*$ (dashed lines) as functions of the stiffness parameter $n$ for the gLJ fluid with (\textbf{a}) $d=2$ and (\textbf{b}) $d=3$. The~dotted lines represent the quantum-corrected Boyle temperature $\TB$, obtained from Equation~\eqref{22} with $q=5\times 10^{-3}$.\label{fig2}}
\end{figure}

Figure~\ref{fig2} shows $T_0^*$ and $T_1^*$ as functions of $n$ for $d=2$ and $d=3$.
While the classical Boyle temperature $T_0^*$ decreases as the potential becomes stiffer, the~first quantum correction $T_1^*$ increases with $n$.
As a result, quantum effects amplify the decrease of the Boyle temperature with increasing stiffness, as~illustrated by the curves representing $T_0^*-qT_1^*$ with $q=5\times10^{-3}$.
This effect is more pronounced in two-dimensional fluids than in three-dimensional~ones.

\section{Application to Noble~Gases}

In the case of the sLJ model, the~influence of quantum effects on the second virial coefficient can be assessed through the relative deviation, written as follows:
\bal
\label{deltaB}
\delta B_2^*(T)\equiv&\frac{B_2(T)-B^{(0)}(T)}{B_2(T) }=\frac{q}{4}\frac{B_q(T^*)}{B_c(T^*)}+\mathcal{O}(q^2)\nn
=&q\frac{5\Gamma\left(\frac{5}{6}\right)}{32\sqrt{\pi}}\left(8\beta^*\right)^{\frac{5}{6}}\frac{\DD_{\frac{1}{6}}\left(-\sqrt{2\beta^*}\right)
+\DD_{-\frac{11}{6}}\left(-\sqrt{2\beta^*}\right)}{\DD_{\frac{1}{2}}\left(-\sqrt{2\beta^*}\right)}+\mathcal{O}(q^2).
\eal
To first order in the quantum parameter $q$, we note that the relative deviation $\delta B_2^*(T)$ is compactly expressed in terms of the parabolic cylinder functions $\DD_{\frac{1}{2}}$, $\DD_{\frac{1}{6}}$, and~$\DD_{-\frac{11}{6}}$.

As a simple application, we consider the noble gases helium (He), neon (Ne), and~argon (Ar), which can be described by the sLJ potential with the parameter values for $\epsilon$ and $\sigma$ displayed in Table~\ref{tab1} \cite{HCB64}. The~atomic masses, $m$, and~the dimensionless quantum parameter, $q$, defined in Equation~\eqref{21} are also included in Table~\ref{tab1}.

\begin{table}[H]
\caption{Values of $m$, $\epsilon$, $\sigma$, and~$q$ for the noble gases He, Ne, and~Ar.\label{tab1}}
\newcolumntype{C}{>{\centering\arraybackslash}X}
\begin{tabularx}{\textwidth}{CCCCC}
\toprule
\textbf{Gas}	& \boldmath{$m$}~\textbf{(kg)} 	& \boldmath{$\epsilon/k_B$}~\textbf{(K)} & \boldmath{$\sigma$}~\textbf{(m)} &\boldmath{$q$}\\
\midrule
He & $6.646\times 10^{-27} $ &$10.22$ & $2.576\times 10^{-10}$& $0.179$\\
Ne & $3.351\times 10^{-26} $ &$35.70$ & $2.749\times 10^{-10}$& $8.91\times 10^{-3}$ \\
Ar & $6.634\times 10^{-26} $ &$119.8$ & $3.405\times 10^{-10}$&$8.74\times 10^{-4}$ \\
\bottomrule
\end{tabularx}
\end{table}

\begin{figure}[H]
\includegraphics[width=13.8cm]{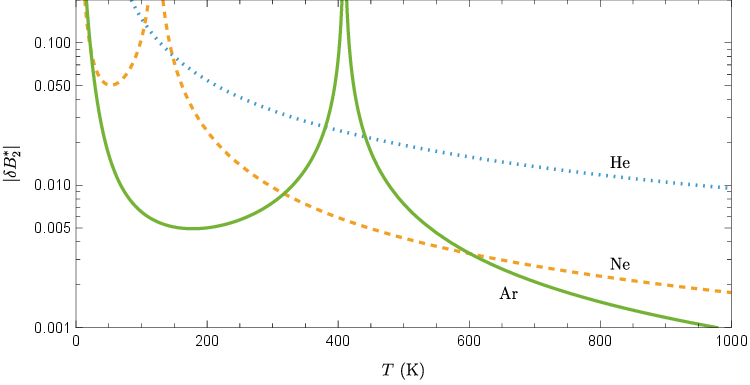}
%
\vspace{-3pt}
\caption{
Absolute value of the relative deviation, $|\delta B_2^*(T)|$, for~helium, neon, and~argon, obtained from Equation~\eqref{deltaB} to first order in $q$.\label{fig3}}
\end{figure}

When comparing $\delta B_2^*$ for  helium, neon, and~argon, one finds that two competing effects are at play. On~the one hand, since $\epsilon_{\text{He}}<\epsilon_{\text{Ne}}<\epsilon_{\text{Ar}}$, a~common fixed temperature $T$ corresponds to  $T^*_{\text{He}}\simeq 11.7\, T^*_{\text{Ar}}>T^*_{\text{Ne}}\simeq 3.36\, T^*_{\text{Ar}}>T^*_{\text{Ar}}$, so that, in~view of Figure~\ref{fig1}b for $n=6$, $B_{q,\text{He}}<B_{q,\text{Ne}}<B_{q,\text{Ar}}$. On~the other hand, $q_{\text{He}}\simeq 204.4 \, q_{\text{Ar}}>q_{\text{Ne}}\simeq 10.2 \,q_{\text{Ar}}>q_{\text{Ar}}$. The~second effect dominates, as~can be observed in Figure~\ref{fig3}.

Note that, because~$B_c(T^*)$ vanishes at the classical Boyle temperature $T_0^*\simeq 3.42$, $|\delta B_2^*(T)|$ diverges at $T=T_0^*\epsilon/k_B$, as~seen in Figure~\ref{fig3} for neon and argon. Apart from this singularity, since quantum effects scale roughly as $(m\epsilon)^{-1}$, helium exhibits the strongest quantum corrections, neon has moderate corrections, and~argon has the weakest quantum effects at any given temperature.
This behavior explains why helium remains liquid at atmospheric pressure down to absolute zero (quantum effects prevent solidification), while argon solidifies at $84$~K and behaves largely classically at typical~temperatures.

\section{Outlook}
\label{sec5}
Although in this work we have focused on the first quantum correction to $B_2$, the~same methodology can be extended to higher-order terms. The~convergence of the quantum expansion in powers of $\hbar^2/m$ depends on the temperature regime and the strength of the interaction. For~systems like helium at very low temperatures, higher-order corrections may become necessary for quantitative~accuracy.

In particular, the~general expression for the second-order correction reads~\cite{K55,KMS55} as follows:
\bal
\label{26}
B^{(2)}(T)=&-\frac{\Omega_d\beta^4}{24}\int_0^\infty \dd r\, r^{d-1} \ee^{-\beta\phi(r)}\left\{\frac{1}{10}\left[\frac{\dd^2\phi(r)}{\dd r^2}\right]^2+\frac{1}{5r^2}\left[\frac{\dd\phi(r)}{\dd r}\right]^2 \right.\nn &\left.+\frac{\beta}{9r}\left[\frac{\dd\phi(r)}{\dd r}\right]^3-\frac{\beta^2}{72}\left[\frac{\dd\phi(r)}{\dd r}\right]^4\right\}.
\eal
Specializing to the gLJ potential, Equation~\eqref{3}, and~introducing the change of variable $r\to t=\sqrt{8\beta^*}(r/\sigma)^{-n}$, one can express $B^{(2)}$ in terms of the parabolic cylinder functions
$\DD_{a-2}(-\sqrt{2\beta^*})$, $\DD_{a-3}(-\sqrt{2\beta^*})$, $\ldots$, $\DD_{a-8}(-\sqrt{2\beta^*})$, with~$a=\frac{d-4}{n}$.
This expression can be further simplified through repeated application of Equation~\eqref{10}.

The systematic nature of our approach---reducing complex integrals to combinations of parabolic cylinder functions---extends naturally to arbitrary orders. This provides a practical computational framework for exploring the convergence properties of the quantum expansion and for determining when higher-order terms become significant. Such analysis would be particularly relevant for light atoms like helium at temperatures below $\sim 50$~K, where the ratio $q B_q(T^*)/4B_c(T^*)$ approaches unity and second-order corrections are no longer~negligible.

\section{Conclusions}
In this paper, we have derived an explicit and compact expression, Equation~\eqref{9}, for~the first quantum correction to the second virial coefficient of a $d$-dimensional fluid composed of particles interacting through the gLJ $(2n,n)$ potential defined in Equation~\eqref{3}.
As in the classical case, Equation~\eqref{6}, the~first quantum correction has been conveniently expressed in terms of parabolic cylinder functions.
For the particular case of the sLJ fluid ($d=3$, $n=6$), the~expression obtained here for $B_q$ [see Equation~\eqref{16}] is considerably more concise than the combination of Equations~\eqref{7} and~\eqref{8} reported previously~\cite{ZGCPBEHANLL25,M66}.

An additional asset of the present results is that they allow one to explore the combined influence of dimensionality and stiffness on the quantum correction $B_q$.
From Equation~\eqref{9}, it follows that the ratio $B_q/n$ depends on $d$ and $n$ only through the combination $(d-2)/n$.
This implies that, at~a given reduced temperature $T^*$, the~value of $B_q/n$ for a $d$-dimensional fluid ($d>3$) with stiffness $n$ is identical to that of a three-dimensional fluid with an effective stiffness $n_{\text{eff}}=n/(d-2)$.
In contrast, for~two-dimensional fluids, $B_q/n$ is independent of $n$ and is given by the particularly simple expression of Equation~\eqref{20}, which is also applicable to any $d$ in the limit $n\to\infty$.

The knowledge of $B_q$ has enabled us to derive the first quantum correction to the Boyle temperature [see Equation~\eqref{25}].
As illustrated by Figures~\ref{fig1} and~\ref{fig2}, the~general trend is that the quantum corrections to both the second virial coefficient and the Boyle temperature become more significant as the potential stiffness increases and the system dimensionality~decreases.

We have applied our results to the noble gases helium, neon, and~argon, demonstrating that the relative quantum correction $\delta B_2^*(T)$ decreases significantly (in absolute value) from helium to argon, primarily due to the strong dependence on the quantum parameter $q\propto(m\epsilon)^{-1}$.
This application illustrates the practical utility of our compact expressions for assessing quantum effects in real physical~systems.

In summary, we have obtained a compact and fully explicit expression for the first quantum correction to the second virial coefficient of a $d$-dimensional gLJ fluid, expressed in terms of parabolic cylinder or generalized Hermite functions.
The formulation unifies the treatment of dimensionality and stiffness, provides analytic access to the limiting behaviors, and~naturally yields the quantum correction to the Boyle temperature.
Beyond its intrinsic theoretical interest, the~approach presented here provides a systematic framework for deriving higher-order quantum corrections (as discussed in {Section}~\ref{sec5})
of relevance in quantum and semiclassical fluid theory, and~its application to noble gases demonstrates its utility for understanding quantum effects in real molecular systems.
The compact analytical nature of our results also makes them particularly valuable for pedagogical purposes, providing students and researchers with transparent expressions that reveal the underlying structure of quantum corrections and facilitate the development of physical intuition about quantum effects in~fluids.

\vspace{6pt}
\authorcontributions{Conceptualization, A.S.; methodology, D.P. and A.S.; software, A.S.; validation, D.P. and A.S.; formal analysis, D.P. and A.S.; investigation, D.P. and A.S.;  writing---original draft preparation, A.S.; writing---review and editing, D.P. and A.S.; visualization, A.S.; supervision, A.S.; funding acquisition, A.S. All authors have read and agreed to the published version of the~manuscript.}

\funding{{A.S.} acknowledges financial support from Grant No.~PID2024-156352NB-I00 funded by MCIU/AEI/10.13039/501100011033/FEDER, UE and from Grant No.~GR24022 funded by the Junta de Extremadura (Spain) and by European Regional Development Fund (ERDF) ``A way of making~Europe.''}

\institutionalreview{{~}Not applicable.
}

\dataavailability{The raw data supporting the conclusions of this article will be made available by the authors on request.}

\conflictsofinterest{The authors declare no conflicts of~interest.}

\abbreviations{Abbreviations}{
The following abbreviations are used in this manuscript:\\

\noindent
\begin{tabular}{@{}ll}
gLJ& generalized Lennard-Jones\\
LJ& Lennard-Jones\\
sLJ& standard Lennard-Jones
\end{tabular}
}

\begin{adjustwidth}{-\extralength}{0cm}

\reftitle{References}

 \PublishersNote{}
\end{adjustwidth}
\end{document}